\documentclass[aps,prd,showpacs]{revtex4}
\usepackage{dcolumn}
\usepackage{amsmath,amssymb,setspace,graphicx,subfigure,epsfig}
\usepackage{amsfonts}
\usepackage{latexsym}
\usepackage{epsfig}
\usepackage{color}
\usepackage{nicefrac}

\newcommand{\dis}[1]{\begin{equation}\begin{split}#1\end{split}\end{equation}}

\newcommand{\be}{\begin{equation}}
\newcommand{\ee}{\end{equation}}

\newcommand{\Mp}{M_P}

\newcommand{\vp}{\varphi}
\newcommand{\vps}{\varphi_*}
\newcommand{\chis}{\chi_*}

\newcommand{\epstar}{\epsilon_\vp^*}

\newcommand{\ecstar}{\epsilon_\chi^*}

\newcommand{\fnlf}{ f^{(4)}_{\rm NL} }

\newcommand{\fnl}{f_{\rm NL}}

\newcommand{\calA}{{\cal A}}

\newcommand{\calF}{{\cal F}}
\newcommand{\calP}{{\cal P}}

\newcommand{\calG}{{\cal G}}
\def\ces{\cos^2\theta^e}
\def\cis{\cos^2\theta^*}

\def\ses{\sin^2\theta^e}
\def\sis{\sin^2\theta^*}
\def\etap{\eta_{\varphi\varphi}}
\def\etac{\eta_{\chi\chi}}

\def\epspi{\epsilon_{\varphi}^*}
\def\epsce{\epsilon_{\chi}^e}

\def\bea{\begin{eqnarray}}
\def\eea{\end{eqnarray}}

\def\bkone{{\bf k_1}}
\def\bktwo{{\bf k_2}}

\def\picube{(2\pi)^3}
\newcommand{\sdelta}[1]{\!\delta^{\,3}(\mathbf{#1})}

\begin{document}
\title{Conditions for large non-Gaussianity in two-field slow-roll inflation}
\author{Christian T. Byrnes}
\email{C.Byrnes@thphys.uni-heidelberg.de}
\affiliation{Institut f\"ur Theoretische Physik, Universit\"at Heidelberg, Philosophenweg 16, 69120
Heidelberg, Germany}
\author{Ki-Young Choi}
\email{kiyoung.choi@uam.es}
\affiliation{Departamento de F\'{\i}sica Te\'{o}rica C-XI,
        Universidad Aut\'{o}noma de Madrid, Cantoblanco,
        28049 Madrid, Spain}
\affiliation{Instituto de F\'{\i}sica Te\'{o}rica UAM/CSIC,
        Universidad Aut\'{o}noma de Madrid, Cantoblanco,
        28049 Madrid, Spain}
\author{Lisa M.H. Hall}
\email{lisa.hall@sheffield.ac.uk}
\affiliation{Department of Applied Mathematics, University of Sheffield, Sheffield, S3 7RH, UK}


\hfill HD-THEP-08-16, FTUAM 08/8, IFT-UAM/CSIC-08-42

%\date{\today}

\begin{abstract}
We study the level of primordial non-Gaussianity in slow-roll two-field
inflation. Using
an analytic formula for the nonlinear parameter $\fnl$ in the case of a
sum or product
separable potential, we find that it is possible to generate significant
non-Gaussianity
even during slow-roll inflation with Gaussian perturbations at Hubble
exit. In this paper
we give the general conditions to obtain large non-Gaussianity and
calculate the level of
fine-tuning required to obtain this. We present explicit models in which
the
non-Gaussianity at the end of inflation can exceed the current
observational bound
of $|\fnl|\lesssim100$.
\end{abstract}

\maketitle
%%%%%%%%%%%%%%%%%%%%%%%%%%%%%%%%%%%%%%%%%%%%%%

%\begin{document}

\section{Introduction}

Standard slow-roll single-field inflation generates a quasi-scale invariant spectrum of Gaussian,
adiabatic perturbations. While this is in agreement with current data, string theory and SUSY
generically contain many scalar fields and it is important to test how these extra fields may change the
predictions of the simplest models. In particular, the simplest models predict a level of
non-Gaussianity which is too small to observe in the foreseeable future
\cite{Maldacena:2002vr,non-gaussian1,Seery:2005gb}, so any detection would be of great significance.

In this paper we focus on the possibility to obtain a large level of non-Gaussianity
during slow-roll inflation and derive general conditions for when it may be large.
We present simple, explicit two-field models which can saturate the observational bound.
Although there are  many models of inflation which generate a large
non-Gaussianity~\cite{Alabidi:2006hg,Bartolo:2001cw,Enqvist:2004ey,Jokinen:2005by,Chen,cline,
Misra:2008tx}, few
of them do so during inflation except by breaking slow-roll, e.g.~with a kink in the
potential \cite{Chen:2006xjb}  or having a non-standard kinetic term, e.g.~DBI inflation
\cite{DBI}. Other widely considered methods to generate a large non-Gaussianity include
the curvaton scenario \cite{curvaton}, modulated reheating \cite{modulatedreheating} or
an inhomogeneous end to inflation
\cite{Bernardeau:2002jf,endofinflation,Alabidi:endinf,Sasaki:2008uc,Dutta:2008if}. All these
scenarios
require an additional light scalar field that doesn't affect the dynamics during
inflation but that becomes important either at the end of inflation or later.

Several authors have considered the possibility of generating non-Gaussianity during inflation for
a separable potential. In particular Vernizzi and Wands calculated a general formula for the non
linearity parameter $f_{NL}$ which parameterises the bispectrum in the case of a sum separable
potential \cite{Vernizzi:2006ve}, this was later extended to the case of a product separable
potential including non-canonical kinetic terms~\cite{Choi:2007su} and a sum separable potential for an arbitrary number of fields
\cite{Battefeld:2006sz}. This has been further extended to the trispectrum (4-point function)
\cite{Seery:2006js}. For other approaches, see for example
\cite{Rigopoulos:2005us,Yokoyama:2007uu}.

Rather less work, however, has been done on analysing the formulas which have been
calculated and are present in the literature. The results are rather long and depend on many parameters.  In general they appear to
give a slow-roll suppressed non-Gaussianity, subject to several simplifying assumptions that are
additional to the slow-roll assumption and that are not always valid. Here we carefully consider
several explicit models of two-field inflation and scan the parameter space more generally
than has 
been done before. We show that it is possible to generate an extremely large non-Gaussianity during
slow-roll inflation even when the field perturbations at Hubble exit are Gaussian. However this is
only possible for specific values of the model parameters and initial conditions. We consider under which conditions a general model can generate a large non-Gaussianity and show
that this always requires some fine tuning of initial conditions.

Although previous papers have shown that it is possible to generate a narrow ``spike" of large
non-Gaussianity while the inflaton trajectory turns a corner, the non-Gaussianity decays again very
quickly after the corner \cite{Vernizzi:2006ve,Rigopoulos:2005us}. This spike of non-Gaussianity is
associated with a temporary ``jump" in the slow roll parameters \cite{Yokoyama:2007uu}. In the
models we consider the non-Gaussianity is large over many $e$-foldings of inflation and all of the
slow roll parameters remain smaller than unity.

The paper is organised as follows: First we define relevant quantities and introduce some notation (Section \ref{sec:theory}).
Then, in the following two sections we derive the general conditions to generate a large non-Gaussianity
during inflation, in Section~\ref{Section_product} for a product separable potential and in
Section~\ref{Section_sum} for a sum separable potential.
In Section~\ref{Section_examples} we give specific examples for product and
sum potentials which can generate large non-Gaussianity. This includes a two-field model of
hybrid inflation in Section~\ref{sec:hybrid}. In Section~\ref{Section_noncanonical} we extend the
previous results to a generalised action with non-canonical kinetic terms.
We conclude in Section~\ref{Section_conclusion}.

%\note{When finished, check this summary:}
%The paper is organised as follows:
%In the next Section we summarise background we use.
%Then we derive the general condition to give
%large non-Gaussianity for the product separable potential in
%Section~\ref{Section_product} and for the sum separable potential in
%Section~\ref{Section_sum}.
%In Section~\ref{Section_examples} we give the specific examples for product and
%sum potentials which give large non-Gaussianity.
%In Section~\ref{Section_noncanonical} we extend the previous results to the
%potentials with non-canonical kinetic terms.
%We conclude in Section~\ref{Section_conclusion}.
%Useful formulae are given in the Appendices.

\section{Background Theory}
\label{sec:theory}

In this paper, we consider an inflationary epoch driven by two scalar fields, whose
dynamics are governed by the action \dis{ S=\int d^4 x
\sqrt{-g}\left[\Mp^2\frac{R}{2}-\frac12g^{\mu\nu}\partial_\mu\vp\partial_\nu\vp
-\frac12g^{\mu\nu}\partial_{\mu}\chi\partial_\nu\chi-W(\vp,\chi) \right]. \label{action}}
Here $\Mp=1/\sqrt{8\pi G}$ is the reduced Planck mass. We consider slow-roll inflation, during
which all of the slow-roll parameters defined below are less than unity.
\bea\label{srdefinition}
&&\epsilon_\vp
=\frac{\Mp^2}{2}\left(\frac{W_\vp}{W}\right)^2,
\qquad
\epsilon_\chi
=\frac{\Mp^2}{2}\left(\frac{W_\chi}{W}\right)^2, \qquad \epsilon = \epsilon_\vp + \epsilon_\chi, \nonumber \\
&&\eta_{\vp\vp}=\Mp^2\frac{W_{\vp\vp}}{W},\qquad
\eta_{\vp\chi}=\Mp^2\frac{W_{\vp\chi}}{W},\qquad
\eta_{\chi\chi}=\Mp^2\frac{W_{\chi\chi}}{W}.
\eea

While it is physically interesting to consider the slow-roll parameters, it will be useful to use
the definition of angles along or perpendicular to the background trajectory of the two
inflationary fields \cite{Gordon:2000hv}:
\bea
\cos\theta = \frac{\dot\vp}{\sqrt{\dot\vp^{2}+\dot\chi^{2}}}, \qquad
\sin\theta = \frac{\dot\chi}{\sqrt{\dot\vp^{2}+\dot\chi^{2}}},
\eea
such that, in the slow roll approximation,
\bea
\frac{\epsilon_{\vp}}{\epsilon} = \cos^{2}\theta, \qquad
\frac{\epsilon_{\chi}}{\epsilon} = \sin^{2}\theta.
\eea

Throughout this paper we use formulae given in \cite{Choi:2007su} and we refer the reader to this
paper and references therein for more details. In summary we define several observable quantities
in terms of the primordial curvature perturbation $\zeta$, which may be calculated using the
$\delta$N formalism \cite{starob85,ss1,Sasaki:1998ug,lms,lr}. The number of
$e$--foldings, $N$, given by
\be N=\int^{t_{\rm{fin}}}_{t_*} H(t)dt,
\ee
is evaluated from an initial flat hypersurface to a final uniform density hypersurface. The
perturbation in the number of $e$--foldings, $\delta N$, is the difference between the curvature
perturbations on the initial and final hypersurfaces. In this paper we take the initial time to be
Hubble exit during inflation, denoted by $t_*$.
%, and the final time to be the end of inflation. 
In this paper, we consider the final time to be a uniform density hypersurface just before the end of inflation, as in \cite{Vernizzi:2006ve,Choi:2007su}.  We note that this differs from the hybrid scenario of \cite{Sasaki:2008uc}, in which the final hypersurface is determined by a separate waterfall field.
The curvature perturbation is given by \cite{lr}
\be\label{deltaN} \zeta=\delta N= \sum_I
N_{,I}\delta\vp_{I*}+\frac12\sum_{IJ}N_{,IJ}\delta\vp_{I*}\delta\vp_{J*}+\cdots\,, \ee
where $N,_I=\partial N/(\partial \vp^I_*)$ and the index $I$ runs over all of the fields.
We will consider the
power spectrum and bispectrum defined (in Fourier space) by
\begin{eqnarray}\label{powerspectrumdefn} \langle\zeta_{\bkone}\zeta_{\bktwo}\rangle &\equiv&
\picube\,
\sdelta{\bkone+\bktwo}\frac{2\pi^2}{k_1^3}\calP_{\zeta}(k_1) \, , \\
\langle\zeta_{{\mathbf k_1}}\,\zeta_{{\mathbf k_2}}\,
\zeta_{{\mathbf k_3}}\rangle &\equiv& \picube\, \sdelta{{\mathbf
k_1}+{\mathbf k_2}+{\mathbf k_3}} B_\zeta( k_1,k_2,k_3) \,. \end{eqnarray}

From this we can define three quantities of key observational interest, respectively the spectral
index, the tensor-to-scalar ratio and the non-linearity parameter
\bea n_{\zeta}-1&\equiv& \frac{\partial \log\calP_{\zeta}}{\partial\log k}, \\
r&=&\frac{\calP_T}{\calP_{\zeta}}=\frac{8\calP_*}{\Mp^2\calP_{\zeta}}, \\
\fnl&=&\frac56\frac{k_1^3k_2^3k_3^3}{k_1^3+k_2^3+k_3^3}
\frac{B_{\zeta}(k_1,k_2,k_3)}{4\pi^4\calP_{\zeta}^2}. \label{fnldefn} \eea
Here $\calP_*$ is the
power spectrum of the scalar field fluctuations and $\calP_T=8\calP_*=8H_*^2/(4\pi^2\Mp^2)$ is the
power spectrum of the tensor fluctuations. 
Both the spectra are calculated at the end of inflation and we ignore any evolution after this time.  For full evolution after inflation, see~\cite{Wands:2002bn,Choi:2007su}.
As defined above, $\fnl$ is shape
dependent, but it has been shown that the shape dependent part is much less than unity
\cite{Lyth:2005qj,Vernizzi:2006ve}. The
ideal CMB experiment is only expected to reach a precision of $\fnl$ around unity
\cite{Komatsu:2001rj}, so we will
calculate the shape independent part of $\fnl$, denoted by $\fnlf$ in
\cite{Vernizzi:2006ve,Choi:2007su}. Whenever the non-Gaussianity is large, $|\fnl|>1$, we can
associate $\fnlf\simeq\fnl$. This ($k$ independent) part of $\fnl$ and the spectral index can be
calculated by the $\delta
N$ formalism,
\bea
{\cal P}_\zeta&=&\sum_I N_{,I}^2 {\cal P}_*,\label{spectrum} \\
n_{\zeta}-1&=& -2\epsilon^* +
\frac{2}{H}\frac{\sum_{IJ}\dot{\vp}_J N_{,JI}N_{,I}}{\sum_K N_{,K}^2},\label{index}
\\
\fnlf&=&\frac56 \frac{\sum_{IJ}N_{,IJ}N_{,I}N_{,J}}{\left(\sum_I N_{,I}^2\right)^2}.
\eea
There is no universal agreement over the sign of $\fnl$ in the literature. We use the opposite sign
convention to \cite{Vernizzi:2006ve,Choi:2007su}, in
order to be in agreement with the WMAP sign convention \cite{Komatsu:2008hk}.
The latest observations from 5 years of WMAP data are
\bea n_{\zeta}&=&0.96^{+0.014}_{-0.013},\qquad (\mathrm{assuming}\,\,r=0), \\
r&<&0.2\,\,\,\,(95\%\,\,\rm{CL}), \\ -9&<&\fnl^{\textrm{local}}<111
\,\,\,\,(95\%\,\,\rm{CL}). \eea
The ``local" bound on $\fnl$ is based on the definition
$\zeta=\zeta_G+3\fnl\zeta_G^2/5$, where $\zeta_G$ is the Gaussian part of $\zeta$
\cite{Komatsu:2001rj,Komatsu:2008hk}. When more than one field direction during inflation
contributes to $\zeta$ in Eqn.~(\ref{deltaN}) then this definition of $\fnl$ is not equivalent
to the very commonly used definition Eqn.~(\ref{fnldefn}), even though they have the same
shape dependence (see e.g.~\cite{Byrnes:2006vq}). The bound on $\fnl^{\textrm{local}}$
may therefore not be applicable to the $\fnl$ that we calculate.
In this paper we will
focus on the non-linearity parameter $\fnl$. Constraints on this parameter are expected
to improve by nearly an order of magnitude with the PLANCK satellite \cite{planck}.

%%%%%%%%%%%%%%%%%%%%%
\section{Large non-Gaussianity in Product Potentials, $W(\vp,\chi)=U(\vp)V(\chi)$}\label{Section_product}
We follow the terminology of \cite{Vernizzi:2006ve} and use the definitions, \be
u\equiv\frac{\epsilon_\vp^e}{\epsilon^e}= \cos^2\theta^e, \quad
v\equiv\frac{\epsilon_\chi^e}{\epsilon^e} = \sin^2\theta^e. \ee We will use a sub-
or superscript ``*'' to denote values evaluated at the time of Hubble exit and a
subscripted ``e'' will denote the time at $t_e$. From Eqns.~(\ref{spectrum})
and~(\ref{index}), the power spectrum and spectral index for a product potential with
canonical kinetic terms are \dis{ {\cal P}_\zeta = \frac{W_*}{24\pi^2\Mp^4
}\left(\frac{u^2}{\epstar} + \frac{v^2}{\ecstar}\right),\label{spectrum_p} } \dis{
n_\zeta-1=-2\epsilon^* -4\frac{u^2\left(1-\frac{\eta_{\vp\vp}^*}{2\epstar}\right) +
v^2\left(1-\frac{\eta_{\chi\chi}^*}{2\ecstar}\right)}{u^2/\epstar +
v^2/\ecstar}\label{index_p}. } The non-linear parameter $\fnlf$
becomes~\cite{Choi:2007su}
\begin{eqnarray}
  f^{(4)}_{\rm NL}&=& \frac{5}{6}
\frac{2}{\left( \frac{u^2}{\epsilon_\vp^*}
+ \frac{v^2}{\epsilon_\chi^*} \right)^2}
\left[
\frac{u^3}{\epsilon^*_\vp}
\left(1
 - \frac{\eta^*_{\vp\vp}}{2 \epsilon^*_\vp}
\right)
+ \frac{v^3}{\epsilon^*_\chi}
\left(1
 - \frac{\eta^*_{\chi\chi}}{2 \epsilon^*_\chi }
\right)
- \left( \frac{u}{{\epsilon^*_\vp}}
- \frac{v}{{\epsilon^*_\chi}}
\right)^2 \calA_P
\right]
,
\label{fNL4product}
\end{eqnarray}
where
\begin{eqnarray}
\hat\eta &\equiv&
\frac{\epsilon_\chi\eta_{\vp\vp}+\epsilon_\vp\eta_{\chi\chi}}{\epsilon}, \\
\calA_P&\equiv&-\frac{\epsilon^e_\vp \epsilon^e_\chi}{(\epsilon^e)^2}
\left[
\hat\eta^e- 4 \frac{\epsilon^e_\vp \epsilon^e_\chi}{\epsilon^e}
\right].
\end{eqnarray}
The slow-roll parameters, defined by Eq.~(\ref{srdefinition}), are
\dis{
\epsilon_\vp
=\frac{\Mp^2}{2}\left(\frac{U_\vp}{U}\right)^2=\epsilon \cos^2\theta,\qquad
\epsilon_\chi
=\frac{\Mp^2}{2}\left(\frac{V_\chi}{V}\right)^2=\epsilon \sin^2\theta,
}
and
\dis{
\eta_{\vp\vp}
=\Mp^2\frac{U_{\vp\vp}}{U},\qquad
\eta_{\vp\chi}
=\Mp^2\frac{U_{\vp}V_{\chi}}{W},\qquad
\eta_{\chi\chi}
=\Mp^2\frac{V_{\chi\chi}}{V}.
}
Using a few auxiliary functions (involving only $\theta^*$ and $\theta^e$), we may write
Eqn.~(\ref{fNL4product}) as
\begin{eqnarray}
 f^{(4)}_{\rm NL}&=&
\frac{5}{6}\left[
2j_p(\theta^*,\theta^e)\epsilon^*
-f_p(\theta^*,\theta^e)\eta^*_{\vp\vp}
-g_p(\theta^*,\theta^e)\eta^*_{\chi\chi}
+ 2 h_p(\theta^*,\theta^e,X)
\left(\hat\eta^e
-4\sin^2\theta^e\cos^2\theta^e\epsilon^e\right)
\right]
,
\label{mnexpansion}
\end{eqnarray}
where the auxiliary functions are defined as
\dis{
f_p(\theta^*,\theta^e)&\equiv
\frac{u^3 \sin^4\theta^*}{\left( u^2 \sin^2\theta^* + v^2 \cos^2\theta^* \right)^2}
= \frac{\tan^4\theta^*}{\left(\tan^2\theta^*+\tan^4\theta^e\right)^2\cos^2\theta^e},\\
g_p(\theta^*,\theta^e)&\equiv
\frac{v^3 \cos^4\theta^*}{\left( u^2 \sin^2\theta^* + v^2 \cos^2\theta^* \right)^2}
= \frac{\tan^8\theta^e}{\left(\tan^2\theta^*+\tan^4\theta^e\right)^2\sin^2\theta^e},\\
h_p(\theta^*,\theta^e)&\equiv
\ses\ces\frac{\left(u \sis - v\cis\right)^2}{\left( u^2 \sin^2\theta^* + v^2 \cos^2\theta^* \right)^2}
= \tan^2\theta^e\frac{\left(\tan^2\theta^*-\tan^2\theta^e\right)^2}{\left(\tan^2\theta^*+\tan^4\theta^e\right)^2},\\
j_p(\theta^*,\theta^e)&\equiv
\frac{\left(u^3\sin^4\theta^*\cos^2\theta^* + v^3\sin^2\theta^*\cos^4\theta^*\right)}{\left( u^2 \sin^2\theta^* + v^2 \cos^2\theta^* \right)^2}
= \frac{\sin^2\theta^*}{\cos^2\theta^e}
\frac{\left(\tan^2\theta^*+\tan^6\theta^e\right)}{
\left(\tan^2\theta^*+\tan^4\theta^e\right)^2}. 
}

By analysing the functions $j_p$, $f_p$, $g_p$ and $h_p$ over the range of allowed values for
$\theta^*$ and $\theta^e$, it is possible to locate regions of parameter space which gives large
$f^{(4)}_{\rm NL}$.  We find that the function $j_p(\theta^*,\theta^e)$ satisfies $0\leq j_p\leq 1$ and therefore the
`j'-term can never lead to significant values of $f^{(4)}_{\rm NL}$.  We therefore ignore this term
in the analysis that follows.
\begin{figure}[!t]
\begin{center}
   \includegraphics[width=0.9\textwidth]{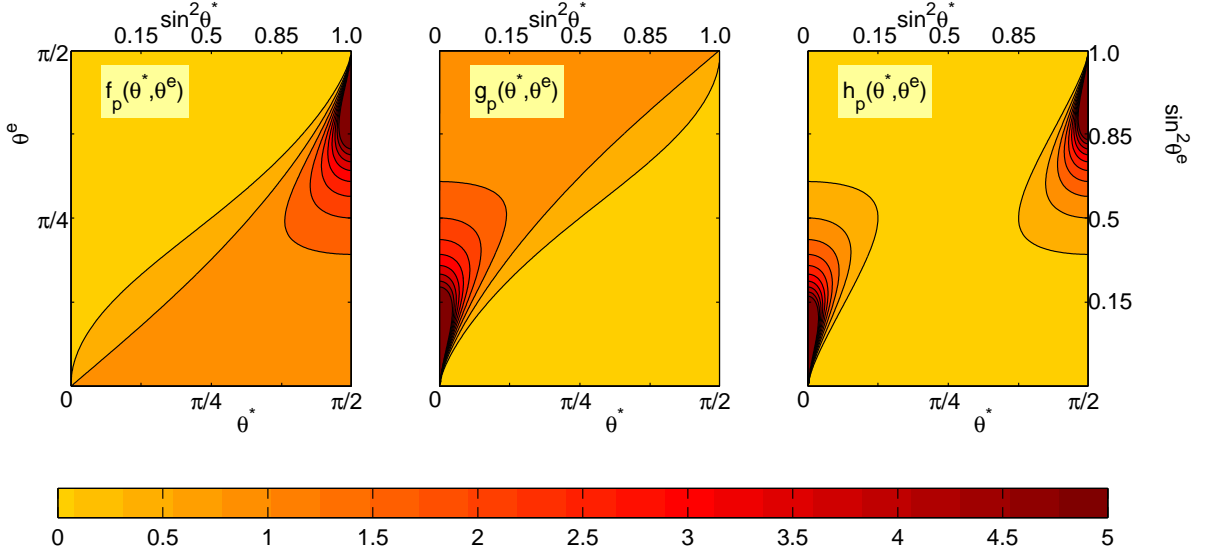}
\end{center}
\caption{The contour plot of the functions, $f_p$, $g_p$ and $h_p$, in the plane of $\theta^*$, and $\theta^e$.  The bottom and left-hand axes show the angles, $\theta^{*}$ and $\theta^{e}$ respectively.  The top and right-hand axes show $\sin^{2}\theta^{*}$ and $\sin^{2}\theta^{e}$.}
\label{mncos}
\end{figure}

The functions $f_p$, $g_p$ and $h_p$ (as functions of $\theta^*$ and $\theta^e$) are plotted in Fig.~\ref{mncos}.  The main point to note is that these prefactors are large (and can be very large) in two regions:
\begin{description}
\item[A] $\cis\ll 1$ and $\ces\ll 1$ ($\epsilon_\chi\gg\epsilon_\vp$).
In this region, $f_p\gg g_p$ and $h_p\sim f_p\sin^2\theta^e$.
\item[B] $\sin^2\theta^*\ll 1$ and $\sin^2\theta^e\ll 1$ ($\epsilon_\vp\gg\epsilon_\chi$).
In this region, $g_p\gg f_p$ and $h_p\sim g_p\cos^2\theta^e$.
\end{description}

In both these regions, these prefactors can account for a significant level of $f^{(4)}_{\rm NL}$, despite the relevant terms in Eqn.~(\ref{mnexpansion}) having slow-roll factors.  Each of these regions describes one of the fields dominating over the other in kinetic energy.  Note, however, that single-field inflation (i.e.~with one field static) will not lead to a large value of non-Gaussianity, since these fractions vanish exactly for $\dot\chi^e=0$ or $\dot\vp^e=0$ ($\theta^e=0$ or $\nicefrac{\pi}{2}$ respectively).

\begin{figure}[!t]
\begin{center}
   \includegraphics[width=0.4\textwidth]{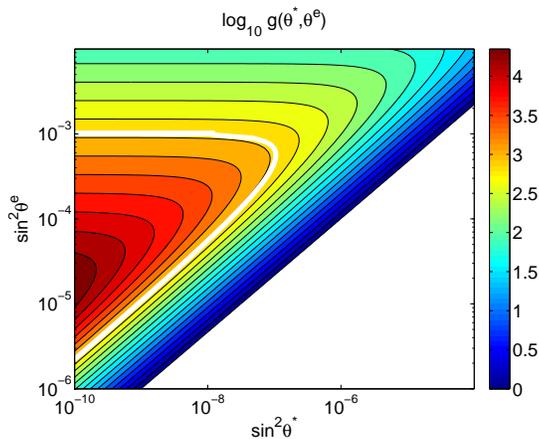}
\end{center}
\caption{A blown-up graph of Region B.
  The conditions for $g_p$ in Eqn.~(\ref{gen:Bconds}) is plotted for ${\cal G}_p=1000$ with a white line. It can be seen that this condition encloses the contour for $g_p$ larger than some constant, ${\cal G}_p$.
}
\label{fclose}
\end{figure}

Due to the symmetry, we shall focus on Region B and explicitly write down full conditions for both Regions A and B in Section~\ref{sec:cond_product}.
To find the condition to give large non-Gaussianity,
we will concentrate on the terms including the functions of $g_{p}(\theta^*,\theta^e)$ and $h_{p}(\theta^*,\theta^e)$,
since the other terms cannot give large non-Gaussianity as discussed above.
The large $\fnlf$ is given by
\bea
\fnlf &\simeq&
\frac{5}{6}g_p(\theta^*,\theta^e)\left[
-\eta^*_{\chi\chi}
+ 2\cos^2\theta^e \left(\hat\eta^e
-4\sin^2\theta^e\cos^2\theta^e\epsilon^e\right)
\right], \nonumber\\
 &\simeq&
\frac{5}{6}
\frac{\sin^6\theta^e}{\left(\sin^2\theta^*+\sin^4\theta^e\right)^2}
\left[
-\eta^*_{\chi\chi} + 2 \eta^e_{\chi\chi}
\right].
\label{gapprox}
\eea
Here we assumed that
$|\eta_{\chi\chi}| \gg \ses \left|\eta_{\vp\vp}^e - 4\epsilon^e \right|$.
For $|\fnl|$ to be larger than unity, the function $g_{p}(\theta^*,\theta^e)$
must be bigger than the inverse of the slow-roll parameters in the square bracket. In this way from Eqn.~(\ref{gapprox}) we find the general way to obtain large non-Gaussianity, $|\fnlf|\gtrsim 1$,
\bea \sis\lesssim\sin^4\theta^e \left(\frac{1}{\sqrt{\ses \calG_p}}-1\right)\, ,
\qquad
 {\cal G}_p= \frac{6}{5}\left| -\eta^*_{\chi\chi} + 2 \eta^e_{\chi\chi} \right|^{-1}.\label{gen:Bconds}
\eea
This bound is shown graphically in Figure \ref{fclose}.

From this condition we can derive some corollaries. Firstly, from $\partial g_p/\partial(\ses)=0$ and the definition of $g_p$, the
initial and final
angle of the trajectory are constrained: 
\bea\label{app:maxcos} \sis<\frac13\left(\frac{3}{4}\right)^4\frac{1}{\calG_p^2}, \quad
\textrm{and}\quad \ses<\frac{1}{\calG_p}, \eea
and secondly we see that $\sin^2\theta$ must typically grow by at least two orders of
magnitude during inflation, since we require
\bea\label{app:cosgrowth} \frac{\ses}{\sis}>4\calG_p. \eea

In this region of large non-Gaussianity, the power spectrum from Eqn.~(\ref{spectrum_p}),
the spectral index from Eqn.~(\ref{index_p}) and the tensor-to-scalar ratio in Region B are:
\dis{ {\cal P}_\zeta
\simeq \frac{W_*}{24\pi^2\Mp^4 \epsilon^*}\left(1+\tilde{r}
\right),\label{spectrum_largeG} } 
\dis{ n_\zeta-1
\simeq -2\epsilon^* +2\frac{-2\epsilon^*+\eta_{\vp\vp}^* + \tilde{r} \eta_{\chi\chi}^*
}{1+ \tilde{r}}, \label{index_largeG} }
\dis{ r\simeq 16\epsilon^* \left(1+\tilde{r}\right)^{-1}, } where \dis{
\tilde{r}=\frac{\sin^4\theta^e}{\sis}>0. }
The spectrum of curvature perturbations can be dominated by the fluctuations in the slow-rolling field
$\vp$ or $\chi$ depending on $\tilde{r}$, which can be smaller or bigger than unity. 
We note, however, that the tensor-to-scalar ratio can only be suppressed
in
multiple field inflation, since the tensor power spectrum is unchanged from the single
field value and $\calP_{\zeta}$ is enhanced. The spectral index is changed by the
presence of $\tilde{r}$, but remains slow roll suppressed.

\subsection{Condition for Large Non-Gaussianity in Product Potentials}
\label{sec:cond_product}
If we summarise,
the general conditions to obtain large $f^{(4)}_{\rm NL}$ can be written simply.  Ultimately, the two-field system must have one field which dominates the evolution throughout.  This corresponds to being in Region A or B (depending on which field dominates).  Hence, one field must dominate the evolution almost fully at the beginning, while the other field gains a relatively large percentage of kinetic energy by the end of inflation.
Specifically, the slow-roll parameters and the value of  $f_{p}$ or $g_{p}$ must be such to give
$|f^{(4)}_{\rm NL}|\gtrsim1$ in Eqn.~(\ref{mnexpansion}).

For Region A ($\cos^2\theta \ll 1$), the condition is
\bea \cis\lesssim\cos^4\theta^e \left(\frac{1}{\sqrt{\ces \calF_p}}-1\right)\, ,
\qquad
 {\cal F}_p= \frac{6}{5}\left| -\eta^*_{\vp\vp} + 2 \eta^e_{\vp\vp} \right|^{-1}\, \quad \textrm{(Region A)}.
\label{Aconds}
\eea
where we have assumed that
$|\eta_{\vp\vp}| \gg \cos^2\theta^e|\eta_{\chi\chi}-4\epsilon^e|$.

For Region B ($\sin^2\theta \ll 1$), the condition is
\bea \sis\lesssim\sin^4\theta^e \left(\frac{1}{\sqrt{\ses \calG_p}}-1\right)\, ,
\qquad
 {\cal G}_p= \frac{6}{5}\left| -\eta^*_{\chi\chi} + 2 \eta^e_{\chi\chi} \right|^{-1}\, \quad \textrm{(Region B)}.
\label{Bconds}
\eea
where we have assumed that
$|\eta_{\chi\chi}| \gg \sin^2\theta^e|\eta_{\vp\vp}-4\epsilon^e|$.

This condition also leads to a large growth of the subdominant slow-roll parameter.  For Region A and B respectively:
\be
\frac{\cos^2\theta^{e}}{\cos^{2}\theta^{*}} = \frac{\epsilon_{\vp}^{e}}{\epsilon^{e}} \frac{\epsilon^{*}}{\epsilon^{*}_{\vp}} \gtrsim 4{\cal F}_p\, \quad \textrm{(Region A)}
, \qquad
\frac{\sin^2\theta^{e}}{\sin^{2}\theta^{*}} = \frac{\epsilon_{\chi}^{e}}{\epsilon^{e}} \frac{\epsilon^{*}}{\epsilon^{*}_{\chi}} \gtrsim 4{\cal G}_p\, \quad \textrm{(Region B)}.
\ee

\subsection{Direct Observation of $f^{(4)}_{\rm NL}$ and Fine-Tuning}
Eqns.~(\ref{Aconds}) and~(\ref{Bconds}) encode the level of fine-tuning required on the initial and final slow-roll parameters to obtain an observable level of non-Gaussianity.
As an example, if observations find $f^{(4)}_{\rm NL}\sim 10$,
for standard order of slow-roll parameters ($\eta\sim0.01$),
one requires $f_p$ or $g_p\sim1000$.
In the case of Region B, this corresponds to
\be
\frac{\epsilon_\chi^*}{\epsilon^*}=\sis \lesssim 10^{-7}, \qquad \textrm{and}
\quad \frac{\epsilon_\chi^e}{\epsilon^e}=\ses \lesssim 10^{-3}.
\ee
This requires very special initial values for the fields.

%%%%%%%%%%%%%%%%%%%%%
\section{Large Non-Gaussianity in Sum Potentials, $W(\vp,\chi)=U(\vp) + V(\chi)$}
%%%%%%%%%%%%%%%%%%%%%
\label{Section_sum}
In this section we find the general condition for large non-Gaussianity  with a sum separable potential,
$W(\vp,\chi)=U(\vp) + V(\chi)$.
Defining
\begin{eqnarray}
u \equiv \frac{U^*+Z^e}{W^*}, \quad \quad
v \equiv \frac{V^*-Z^e}{W^*}, \label{uvsum}
\end{eqnarray}
with
\begin{eqnarray}
Z^e &=& \frac{(V^e {\epsilon^e_\vp} - U^e
{\epsilon^e_\chi})}{\epsilon^e}=V^e\ces -U^e\ses, \label{Z2}
\end{eqnarray}
the power spectrum and spectral index are given by~\cite{Vernizzi:2006ve}:
\dis{
{\cal P}_\zeta = \frac{W_*}{24\pi^2\Mp^4 }\left(\frac{u^2}{\epstar}
+ \frac{v^2}{\ecstar}\right),\label{spectrum_s}
}
\dis{
n_\zeta-1=-2\epsilon^* -4\frac{u\left(1-\frac{\eta_{\vp\vp}^*}{2\epstar}u\right)
+ v\left(1-\frac{\eta_{\chi\chi}^*}{2\ecstar}v\right)}{u^2/\epstar + v^2/\ecstar}\label{index_s}.
}
The nonlinear parameter $\fnlf$ is~\cite{Vernizzi:2006ve}:
\begin{eqnarray}
 f^{(4)}_{\rm NL}= \frac{5}{6}
\frac{2}{\left( \frac{u^2}{\epsilon_\vp^*}
+ \frac{v^2}{\epsilon_\chi^*} \right)^2}
\left[
\frac{u^2}{\epsilon^*_\vp}
\left(1  - \frac{\eta^*_{\vp\vp}}{2 \epsilon^*_\vp} u \right)
+ \frac{v^2}{\epsilon^*_\chi}
\left(1
 - \frac{\eta^*_{\chi\chi}}{2 \epsilon^*_\chi } v \right)
+ \left( \frac{u}{{\epsilon^*_\vp}}
- \frac{v}{{\epsilon^*_\chi}} \right)^2 \calA_S
\right]
,
\label{fNLsum}
\end{eqnarray}
where we define
\begin{eqnarray}
\hat\eta &\equiv&
\frac{(\epsilon_\chi\eta_{\vp\vp}+\epsilon_\vp\eta_{\chi\chi})}{\epsilon}
=\etap\sin^2\theta+\etac\cos^2\theta , \\
\calA_S &\equiv& - \frac{W_e^2}{W_*^2} \frac{\epsilon^e_\vp
\epsilon^e_\chi}{(\epsilon^e)^2} \left[\epsilon^e -\hat\eta^e \right] =  - \frac{W_e^2}{W_*^2}
\ces\ses \left[\epsilon^e -\hat\eta^e \right]
\label{AS}.
\end{eqnarray}

The slow-roll parameters, defined by Eqn.~(\ref{srdefinition}), are
\dis{
\epsilon_\vp
=\frac{\Mp^2}{2}\left(\frac{U_\vp}{U+V}\right)^2=\epsilon \cos^2\theta,\qquad
\epsilon_\chi
=\frac{\Mp^2}{2}\left(\frac{V_\chi}{U+V}\right)^2=\epsilon \sin^2\theta,
}
and
\dis{
\eta_{\vp\vp}=\Mp^2\frac{V_{\vp\vp}}{U+V},\qquad
\eta_{\vp\chi}=0,\qquad
\eta_{\chi\chi}=\Mp^2\frac{V_{\chi\chi}}{U+V},
}

Similar to the analysis of a product potential, it is possible to re-write Eqn.~(\ref{fNLsum}):
\bea
\fnlf
=\frac{5}{6}
\left[
2 j_s \epsilon^*
- f_s \eta^*_{\vp\vp}
- g_s \eta^*_{\chi\chi}
-2 h_s \frac{W_e^2}{W_*^2} \left(\epsilon^e-\hat\eta^e\right)
\right],
\label{fNLsum2}
\eea
where the following auxiliary functions have been used:
\dis{
 f_s(u,\sis)&\equiv\frac{u^3\sin^4\theta^*}{(u^2\sis+v^2\cis)^2},\\
 g_s(u,\sis)&\equiv\frac{v^3\cos^4\theta^*}{(u^2\sis+v^2\cis)^2},\\
 h_s(u,\sis)&\equiv\ses\ces\frac{(u\sis-v\cis)^2}{(u^2\sis+v^2\cis)^2}, \\
 j_s(u,\sis)&\equiv\frac{(u^2\sin^4\theta^*\cos^2\theta^*+v^2\cos^4\theta^*\sin^2\theta^*)}{
(u^2\sis+v^2\cis)^2}.
}

Note that $u+v=1$ and hence these functions depend on just two variables, $u$ and $\theta^*$ (or $v$ and $\theta^*$), and
that $0\leq u\leq 1$ and $0\leq v\leq 1$.  The first term of $\fnlf$ in Eqn.~(\ref{fNLsum2})
is always smaller than unity, since $0\leq j_{s}\leq1$.

Once again there are two regions with potentially large non-Gaussianity:
\begin{description}
\item[C] $u\ll 1$, and $\cis\ll 1$
\item[D]  $v\ll 1$, and $\sis\ll 1$.
\end{description}

We temporarily analyse this second region, Region D, in which $v\ll1, \sis\ll1$ and $g_{s}$ can be large. In this region, we can approximate
\bea\label{g_sapprox}
g_s(v,\sis) \simeq \frac{v^{3}}{(\sis+v^{2})^{2}},
\qquad
 h_s(v,\sis)\simeq  \frac{g_s(v,\sis)\sin^2\theta^e \cos^2\theta^e}{v}.
\eea
In the special type of potential when $V^e\simeq V^*$, such as hybrid inflation, $v\sim\ses W^e/W^*$ and hence
\bea
 h_s(v,\sis)\simeq g_s(v,\sis)\cos^2\theta^e\frac{W_*}{W_e}.
\eea

The function $g_{s}$ has exactly the same form as the function $g_{p}$ for the product potential in the region where it
is large, see Eqn.~(\ref{gapprox}) replacing $\ses$ with $v$. Hence it is
large in exactly the same areas, and for $g_{s}\gtrsim \calG_{s}$ we require
\bea \sis\lesssim v^2 \left(\frac{1}{\sqrt{v\, \calG_s}}-1\right)\, ,
\qquad
 {\cal G}_s= \frac{6}{5}\left| -\eta^*_{\chi\chi} + 2 \eta^e_{\chi\chi} \right|^{-1}\, \quad \textrm{(Region B)}.
\label{gsregD}
\eea
Finally, we can approximate $\fnlf$ in Region D as
\dis{
\fnlf\simeq  \frac{5}{6} g_s
\left[ -\eta^*_{\chi\chi} - 2 \frac{W_e^2}{W_*^2}
\ses\ces \frac{\left(\epsilon^e -\hat\eta^e \right)}{v}  \right],
\label{regDapprox}
}
where we have used $u\simeq 1$ and $\cis\simeq 1$.
We may analyse this equation for the specific case when $V^e\simeq V^*$, which
can be true in many models because $\sis\simeq(V_{\chi}/U_{\vp})^2\ll1$, so at least initially
the potential $V$ is extremely flat.
In this scenario, $v\sim\ses W^e/W^*$ and hence
\dis{\label{regDapproxVconst}
\fnlf\simeq  \frac{5}{6} g_s
\left[ -\eta^*_{\chi\chi} - 2 \frac{W_e}{W_*}
\ces \left(\epsilon^e -\hat\eta^e \right)  \right].
}

Indeed, the form of power spectrum and spectral index in this large non-Gaussianity region, are same as that of product potential,
Eqn.~(\ref{spectrum_largeG}) and~(\ref{index_largeG}).

\subsection{Condition for Large Non-Gaussianity in Sum Potentials}
From the previous section, it is clear that, to get large non-Gaussianity, we need a constraint on the function $g_s$, for Region C,
\bea \cis\lesssim u^2 \left(\frac{1}{\sqrt{u \,\calF_s}}-1\right)\, ,
\qquad
 {\cal F}_s= \frac{6}{5}\left| -\eta^*_{\vp\vp} + 2 \eta^e_{\vp\vp} \right|^{-1}\, \quad \textrm{(Region C)}.
\label{Cconds}
\eea

In the same way, for Region D, we obtain
\bea \sis\lesssim v^2 \left(\frac{1}{\sqrt{v\, \calG_s}}-1\right)\, ,
\qquad
 {\cal G}_s= \frac{6}{5}\left| -\eta^*_{\chi\chi} + 2 \eta^e_{\chi\chi} \right|^{-1}\, \quad \textrm{(Region D)}.
\label{Dconds}
\eea
As in Eqn.~(\ref{app:maxcos}), we find exact upper limits on the parameters:
\bea 
\cis<\frac13\left(\frac{3}{4}\right)^4\frac{1}{\calF_s^2}, \quad u<\frac{1}{\calF_s} \quad \textrm{and} \quad \frac{u}{\cis}>4\calF_p  \quad \textrm{(Region C)} 
\\ 
\sis<\frac13\left(\frac{3}{4}\right)^4\frac{1}{\calG_s^2}, \quad v<\frac{1}{\calG_s} \quad \textrm{and} \quad \frac{v}{\sis}>4\calG_p. \quad \textrm{(Region D)}
\eea
%

%%%%%%%%%%%%%%%%%%%%%
\section{Examples}
\label{Section_examples}
%%%%%%%%%%%%%%%%%%%%%
In this section we give specific examples, two of which can give large
non-Gaussianity and one which cannot. We note that, in all the following examples, we
have one parameter that acts as a normalisation factor in $W_*$. This parameter can be fixed to normalise the amplitude of the perturbations.

%%%%%%%%%%%%%%%%%%%%%
\subsection{Quadratic times exponential potential, $W(\vp,\chi)=\frac12 e^{-\lambda\vp^2/\Mp^2}
m^2\chi^2$}
%%%%%%%%%%%%%%%%%%%%%

We consider a product potential, with $U(\vp)=e^{-\lambda\vp^2/\Mp^2}$ and $V(\chi) =
m^2\chi^2/2$, for which the slow-roll parameters are: \dis{ \epsilon_\vp = 2\lambda
^2\frac{\vp^2}{\Mp^2}, \qquad \eta_{\vp\vp}=-2\lambda
+4\lambda^2\frac{\vp^2}{\Mp^2},\qquad \epsilon_\chi=
\eta_{\chi\chi}=\frac{2\Mp^2}{\chi^2},\qquad \eta_{\vp\chi}=-\frac{4\lambda\,\vp}{\chi}. \label{ex1_slow}
} We consider $\lambda>0$ and slow-roll throughout inflation, so that the $\vp$-field increases during inflation
while the $\chi$-field decreases. In this slow-roll limit, the exponential function can be expanded so that the potential is dominated by the quadratic potential of $\chi$ field, $\frac{1}{2}m^2\chi^2$, and corrected by the integration between $\chi$ and $\vp$ field, $-\frac12\lambda m^2\vp^2\chi^2$, which might be important for preheating~\cite{Kofman:1994rk,Kofman:1997yn}.
Since inflation ends when the $\chi$ field rolls close to its minimum, i.e.~when
$\epsilon=1\simeq\epsilon_\chi$, the only way to generate large non-Gaussianity is to
start with small field of $\vp$ (which corresponds to Region A).

For this potential, the slow-roll solutions for $\vp$ and $\chi$ lead to
\dis{
\vp=\vps e^{2\lambda N} \qquad \chi^2= \chis^2 - 4N\Mp^2.
}
Since  $\lambda^2\vp^2/\Mp^2<1$,  we find $\eta^*_{\vp\vp}\sim\eta^e_{\vp\vp}\sim -2\lambda$.
From Eqn.~(\ref{Aconds}), the constraints on the initial values of the field $\vp$ can be obtained:
\begin{eqnarray}
\frac{\Mp^2\chi^4_*e^{-12\lambda N_e}}{2\lambda^3\chi_e^6}
&\lesssim& \frac{\vp_*^2}{\Mp^2} \lesssim
\frac{2\Mp^2e^{-4\lambda N_e}}{\lambda\chi_e^2},
\end{eqnarray}
where $N_e\simeq (\chi_*^2-\chi_e^2)/4\Mp^2$ and $\chi_e^2\simeq 2$.
For the specific values of $\chi_*=16\Mp$ and $\lambda = \{0.03,0.04,0.05\}$, the
constraints lead to
\begin{eqnarray}
\{ 0.134,2\times 10^{-3},3\times 10^{-5}\}  &\lesssim& \frac{|\vp_*|}{\Mp} \lesssim \{
0.128,0.031,0.0078\}. \nonumber
\end{eqnarray}
In this range, the non linearity parameter is well approximated as
\dis{
\fnlf\simeq
 \frac56 \frac{\cos^6\theta^e}{(\cis+\cos^4\theta^e)^{2}}\etap.\label{fnlA}
}

\begin{figure}[!t]
\begin{center}
   \includegraphics[width=0.8\textwidth]{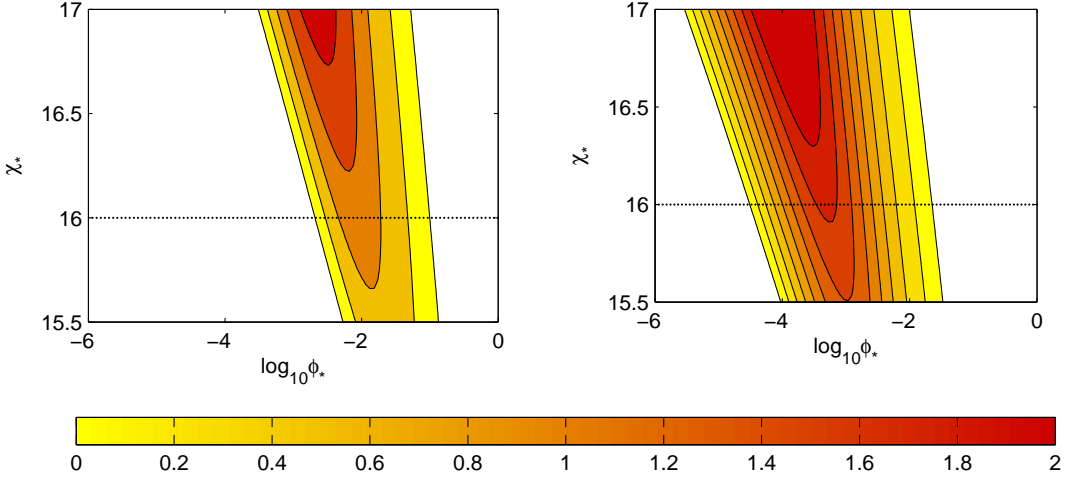}
\end{center}
 \caption{The contour plot of $\log_{10}|\fnlf|$ for Example A, $W(\vp,\chi)=\frac12 e^{-\lambda\vp^2/\Mp^2}
m^2\chi^2$. Here we used $\lambda=0.04$ (left) and $0.05$ (right). 
White regions indicate when $\vert\fnlf\vert<1$ and the dark
centre-most region indicates $|\fnlf|>100$.}
\label{fnlcont_exp_quad}
\end{figure}

The contour plots of $\fnlf$ for two values of $\lambda$ ($\lambda=0.04,0.05$) are given in Fig.~\ref{fnlcont_exp_quad} using the full formula Eqn.~(\ref{fNLsum}), which is almost same as Eqn.~(\ref{fnlA}). Here it can be seen that $\fnlf$ can be very large when $\vp^*\sim 10^{-3}\Mp$ ($\lambda=0.04$) or $\vp^*\sim 10^{-4}\Mp$ ($\lambda=0.05$).  We note that the analytic constraints differ slightly from the plots, due to the value of $\eta_{\vp\vp}$ used.  In the analytics, we assume $\eta_{\vp\vp}\sim-2\lambda$, but use the exact form (Eqn.~(\ref{ex1_slow})) in the plots.

\begin{figure}[!t]
  \begin{center}
   \includegraphics[width=1.0\textwidth]{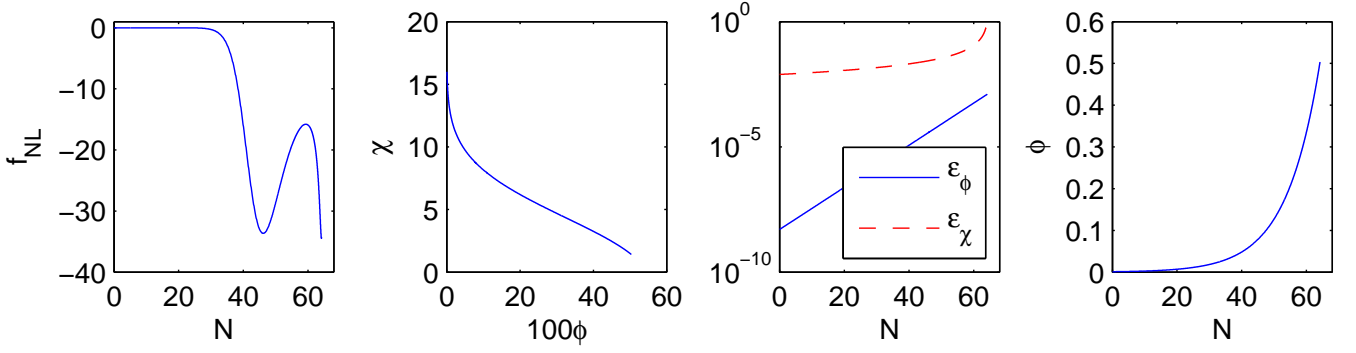}
  \end{center}
 \caption{The analytic evolution of $\fnlf$ and $\epsilon_\chi$
and $\epsilon_\vp$ for Example A.  The evolution of both fields, $\vp$ and $\chi$ are also shown.
We used $\lambda = 0.05$ and $\vp_*=10^{-3}\Mp$, $\chi_*= 16\Mp$.  We numerically solved the full equations of motion until $\epsilon=1$.
}
\label{evol_fnl}
\end{figure}

The time evolution of $f^{(4)}_{\rm NL}$ from the $\delta N$ formula is given in Fig.~\ref{evol_fnl}, where the final point is identified as some time during inflation.
This method is also used in Vernizzi and Wands~\cite{Vernizzi:2006ve} for analytic evolution.
We can see that $|\fnlf|$ increases sharply around $N=30$. This can be understood
from the evolution of two fields: from this time, the velocity of $\vp$ (or $\cos^2\theta$) increases (as seen in the third and fourth plot of Fig.~\ref{evol_fnl})
to satisfy the condition Eqn.~(\ref{Aconds}) to obtain large non-Gaussianity.
Then $\fnlf$ begins to decrease around $N=45$ as $\cos^4\theta^e$ becomes bigger than $\cos^2\theta^*$.
Note that, around the region for large $|\fnl|$, the curvature perturbation of
$\vp$-field becomes comparable to, and then dominates, that of the $\chi$-field.
Around $N=58$, the $\chi$-field begins to roll so fast that $\cos^2\theta$
decreases and $|\fnlf|$ increases once more.  However during this fast-roll stage, slow-roll breaks down and our analytic
formula is not valid any more. Thus we calculate final values when $\epsilon=1$.

We note, for reference, that in this regime of large non-Gaussianity, the background trajectory must be at almost entirely in the one-field direction.  
We therefore expect reheating to take place in almost the same way as it would in normal chaotic inflation.

%%%%%%%%%%%%%%%%%%%%%
\subsection{Quadratic product potential}
%%%%%%%%%%%%%%%%%%%%%

In this section we show that a direct coupling between $\vp$ and $\chi$, leaving to the product potential 
\bea W=\frac{\lambda}{2}\vp^2\chi^2, \eea
cannot generate a large non-Gaussianity during inflation. From the equation $N=\int H dt$ it
follows that the fields evolve according to
\bea \vp_e^2=\vp_*^2-4N\Mp^2 \eea
and similarly for $\chi$. Hence we have in Region B, where $\sis\ll1$
and $4N\Mp^2 < \vp_*^2 \ll \chi_*^2$,
\bea \ses\simeq
\frac{\epsce}{\epsilon^e}=\frac{\vp^2_*-4N\Mp^2}{\chi^2_*-4N\Mp^2}<\frac{\vp^2_*}{\chi^2_*}=\sis.
\eea
Hence $\sin^2\theta$ decreases during inflation and this model cannot generate $|\fnl|>1$.

%%%%%%%%%%%%%%%%%%%%%
\subsection{Hybrid inflation}\label{sec:hybrid}
%%%%%%%%%%%%%%%%%%%%%

We consider a model of 2 field hybrid inflation,
\bea\label{Whybrid}
W(\vp,\chi)= W_0\left(1+\alpha\frac{\vp^2}{\Mp^2}+\beta\frac{\chi^2}{\Mp^2}\right), \eea
which is vacuum dominated, i.e.~which satisfies $\left|\alpha\vp^2\right|\ll\Mp^2$ and
$\left|\beta\chi^2\right|\ll\Mp^2$. Here we assume that inflation ends abruptly by
another waterfall field which we don't write down in the potential above. 
We note
that $\fnl$ may change depending on the details of how the waterfall field is coupled to
the two inflaton fields \cite{Bernardeau:2002jf,endofinflation,Sasaki:2008uc}, but in
general this is unlikely to generate a large contribution to $\fnl$ without fine-tuning~\cite{Alabidi:endinf}.

In this regime the slow-roll solutions are \cite{Alabidi:2006hg}, (we can identify
$\eta_\vp=2\alpha$ and
$\eta_\sigma=2\beta$ in the notation of that paper)
\dis{\label{hybridsoln}
\vp= \vp_*e^{-2\alpha N}, \qquad \chi= \chi_*e^{-2\beta N},
}
and the slow-roll parameters are
\dis{\label{hybridsr}
& \eta_{\vp\vp}=2\alpha, \qquad\eta_{\chi\chi}=2\beta,\qquad
\eta_{\vp\chi}=0,\\
&\epsilon_\vp=2\alpha^2 \frac{\vp^2}{\Mp^2}\ll\left|\etap\right|,\qquad
\epsilon_\chi=2\beta^2 \frac{\chi^2}{\Mp^2}\ll\left|\etac\right|.
}
We note that the dominant slow-roll parameters $\etap$ and $\etac$ are constant in this model.

For this case, since $W^e\simeq W^*$, from Eqn.~(\ref{regDapprox}) (or equivalently Eqn.~(\ref{regDapproxVconst})) in Region D, where $\sis\ll1$ and $v\simeq\ses\ll1$, it follows that
\dis{\label{fNLhybrid}
\fnlf\simeq  \frac{5}{6} g_s
\left[ -\eta_{\chi\chi} - 2 \left(\epsilon^e -\hat\eta^e \right)  \right]\simeq \frac56
\frac{\sin^6\theta^e}{(\sis+\sin^4\theta^e)^{2}}\etac.
}
In the above we have used Eqn.~(\ref{g_sapprox}) for $g_s$ in Region D. We hence require the
condition, Eqn.~(\ref{Dconds}), 
\bea \sis\lesssim v^2 \left(\frac{1}{\sqrt{v\, \calG_s}}-1\right)\, ,
\qquad
 {\cal G}_s= \frac{6}{5}\left| \eta_{\chi\chi} \right|^{-1},\,
\eea
and find the exact upper limits on the parameters:
\bea 
\sis<\frac13\left(\frac{5}{6}\right)^2\left(\frac{3}{4}\right)^4\left| \eta_{\chi\chi} \right|^2, \quad
\textrm{and}\quad v=\ses<\frac{5}{6}\left| \eta_{\chi\chi} \right|.
\eea
Noting that in Region D
$\sin\theta=\beta\chi/(\alpha\vp)$, from Eqn.~(\ref{hybridsoln}) we require $N(\alpha-\beta)>1$ so
that $\sin^2\theta$ grows significantly during inflation.

\begin{figure}[!t]
  \begin{center}
   \includegraphics[width=0.8\textwidth]{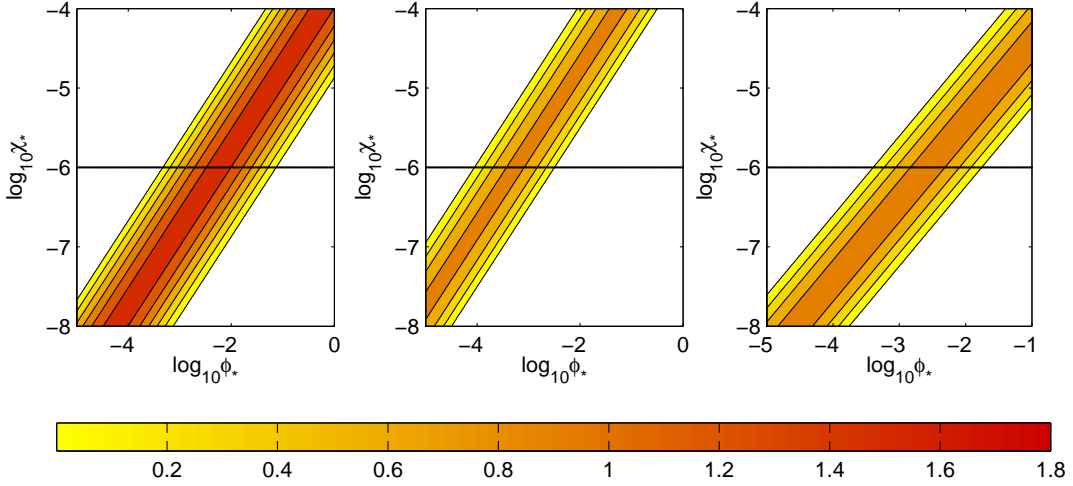}
  \end{center}
 \caption{The contour plot of $\log_{10}\vert\fnlf\vert$ with the sum potential,
$W(\vp,\chi)=V_0(1+\alpha\vp^2+\beta\chi^2)$ (Example C).
The parameters match the examples in Table~\ref{table_hybrid} and the contours for $\log_{10}\vert\fnlf\vert$ are shown.
White regions indicate when $\vert\fnlf\vert<1$.
}
\label{fnlcont_sum}
\end{figure}

In Table~\ref{table_hybrid}, we give some explicit examples of values of $\alpha,\,\beta,\,\vps$ and $\chis$ which
lead to a large non-Gaussianity. Using Eqn.~(\ref{index_s}) we also calculate the
spectral index. 
The contours of $\fnlf$ of this sum potential for a specific choice of $\alpha$ and $\beta$ is given
in Fig.~\ref{fnlcont_sum}
The first example in the Table~\ref{table_hybrid} shows that it is possible to have
$\fnl\simeq50$ and a scale invariant spectrum. The tensor-to-scalar ratio is also
small in this example. 
%
%\newline
%\be
\begin{table}
\begin{tabular}{|cc|cc|c|c|c|}
\hline
$\alpha$ & $\beta$ & $\vps/\Mp$ & $\chis/\Mp$ & $\fnl$ & $n_{\zeta}-1$ & r \\
\hline
  0.018 & -0.018 & 1 & 0.00018 & -42 & 0
& 0.006 \\ 0.04 & 0.005 & 1 & 0.0018 & 9.27 & 0.09 & 0.10  \\ 0.01 & -0.02 & 1 & 0.00037 & -11.1 &
-0.02 & 0.026 \\
\hline
\end{tabular}
\caption{Table showing some initial conditions for the hybrid inflation model that lead to large
levels of non-Gaussianity. The spectral index, calculated from Eqn.~(\ref{index_s}), and the
tensor-to-scalar ratio are also shown.
}
\label{table_hybrid}
\end{table}
%\ee
%

Our formula for $\fnlf$ is also consistent with the calculation of
\cite{Alabidi:2006hg} but our analysis of the parameter space is more extensive and in particular it
allows for very small initial field values. We therefore find a region with much larger
non-Gaussianity than was found in \cite{Alabidi:2006hg}.

As a further check on the algebra, we note that it is also
possible to analyse this model using the formalism of a product potential with
$W(\vp,\chi)=W_0\exp(\alpha\vp^2/\Mp^2)\exp(\beta\chi^2/\Mp^2)$. This is equivalent to the sum
potential
Eqn.~(\ref{Whybrid}) in the limit of vacuum domination ($|\alpha|\vp^2/\Mp^2\ll1$ and
$|\beta|\chi^2/\Mp^2\ll 1$). We have checked that this gives the same
results.
We note from Table~\ref{table_hybrid} that we require a very small value of $\chis$ in Region D. If the
value becomes too small then the motion of the $\chi$ field will become dominated by quantum
fluctuations rather
than the classical drift down the potential, $3H\dot{\chi}\simeq-W_{\chi}=-W_0\beta\chi$,
which we have
assumed.
In order that we can neglect the effect of the quantum fluctuations, the condition we require on the background trajectory is $|\dot{\chi}|\pi/H^2>\sqrt{3/2}$ \cite{Creminelli:2008es}. Using
$\calP_{\zeta}\simeq 10^{-10}$, $H^2\simeq W_0$ and Eqn.~(\ref{spectrum_s}) the condition that the
classical trajectory is valid can be rewritten as
$(1+\tilde{r})\beta^2\chis^2/(\alpha^2\vps^2)>6\calP_{\zeta}$, where
$\tilde{r}=\sin^4\theta^e/\sis$.
We have checked that this is satisfied for all of
the examples given in the Table~\ref{table_hybrid}, but it does provide a significant constraint on the total parameter space.

Very recently Cogollo {\it et al.~}have calculated the effect of the loop correction to the
primordial power spectrum and bispectrum \cite{Cogollo:2008bi}. This loop correction arises from
taking into account the contribution to the power spectrum and bispectrum arising from terms in the
$\delta N$ expansion which are non-leading in the expansion of the field perturbation $\delta\vp$,
see e.g.~\cite{Byrnes:2007tm}. However they can still be significant if the coefficient to the term
given by the derivative of $N$ is extremely large, e.g.~\cite{Boubekeur:2005fj}. These ``higher
order" terms are usually neglected, but \cite{Cogollo:2008bi} has shown the first explicit example
of an inflation model where they cannot be neglected. They consider a 2-field hybrid model with the
same potential as Eqn.~(\ref{Whybrid}) in the special case of an unstable straight trajectory along one
of the axes, with $\alpha$ and $\beta$ both negative. In this case, they find (for certain initial
values) the loop correction is dominant and can generate an observable $\fnl$. We have checked that
for the explicit values given in the table the loop correction does not dominate (under the
assumption that the same loop correction is dominant here as the one in \cite{Cogollo:2008bi}).
However this does provide a further restriction on the allowed parameter space of the ``tree level"
calculation of the model, which is assumed in Eqn.~(\ref{fNLhybrid}). We plan to return to this issue,
and make a more thorough investigation of the hybrid model in a future publication.

%%%%%%%%%%%%%%%%%%%%%
\section{Non-Gaussianity with Non-Canonical Kinetic Terms (Product Potentials)}
\label{Section_noncanonical}
%%%%%%%%%%%%%%%%%%%%%

Generally, the scalar fields need not only couple through the potential, but may also couple kinetically~\cite{Bellio-Wands95,Bellio-Wands96,Starobinsky01,DiMarcoFinelli03}.
In this section, therefore, we consider inflation governed by the following generalised action:
\dis{
S=\int d^4 x \sqrt{-g}\left[\Mp^2\frac{R}{2}-\frac12g^{\mu\nu}\partial_\mu\vp\partial_\nu\vp
-\frac12e^{2b(\vp)}g^{\mu\nu}\partial_{\mu}\chi\partial_\nu\chi-W(\vp,\chi) \right]. }
When the kinetic energies are non-canonical, the local non-Gaussianity is altered from the previous
sections.  In the case of the sum potential, the modifications are non-trivial and are difficult to
analyse.  When the potential is of product form, however, we may  at least partially use the
previous analysis. We therefore concentrate only on a product potential, but our findings relate also to sum potentials.

With a product potential, the local non-Gaussianity is given by \cite{Choi:2007su}:
\begin{eqnarray}
\frac{6}{5}  f^{(4)}_{\rm NL}&=&
\frac{2 e^{-2b_e+2b_*}}{\left( \frac{u^2\alpha^2}{\epsilon_\vp^*}
+ \frac{v^2}{\epsilon_\chi^*} \right)^2}
\left[
\frac{u^3\alpha^3}{\epsilon^*_\vp}
\left(1
 - \frac{\eta^*_{\vp\vp}}{2 \epsilon^*_\vp}
\right)
+ \frac{v^3}{\epsilon^*_\chi}
\left(1
 - \frac{\eta^*_{\chi\chi}}{2 \epsilon^*_\chi }
\right) \right.
\nonumber
\\
&&
\quad\left.
+ \frac12 {\rm sign}\left(b_\vp\right) {\rm sign}\left(\frac{U_\vp}{U}\right)
\frac{vu^2\alpha^2}{(\epsilon_\vp^*)^2} \sqrt{\epsilon^*_b\epsilon^*_\vp}
- \left( \frac{u\alpha}{{\epsilon^*_\vp}}
- \frac{v}{{\epsilon^*_\chi}}
\right)^2
e^{2b_e-2b_*} \calA_P
\right],
\label{non-K}
\end{eqnarray}
where
\begin{eqnarray}
\alpha&\equiv& e^{-2b_e+2b_*}\left[1+\frac{\epsilon_\chi^e}{\epsilon_\vp^e}
\left(1-e^{2b_e-2b_*}\right)\right] \\
&=&e^{-X} \left[1+\tan^2\theta^e
\left(1-e^{X}\right)\right], \\
\calA_P&\equiv&-\frac{\epsilon^e_\vp \epsilon^e_\chi}{(\epsilon^e)^2}
\left[
\eta_{ss}^e -
\frac12{\rm sign}\left(b_\vp\right) {\rm sign}\left(\frac{U_\vp}{U}\right)
\frac{(\epsilon^e_\chi)^2}{\epsilon^e} \sqrt{\frac{\epsilon^*_b}{\epsilon^*_\vp}}
- 4 \frac{\epsilon^e_\vp \epsilon^e_\chi}{\epsilon^e}
\right].
\end{eqnarray}

\begin{figure}[!t]
  \begin{center}
   \includegraphics[width=0.95\textwidth]{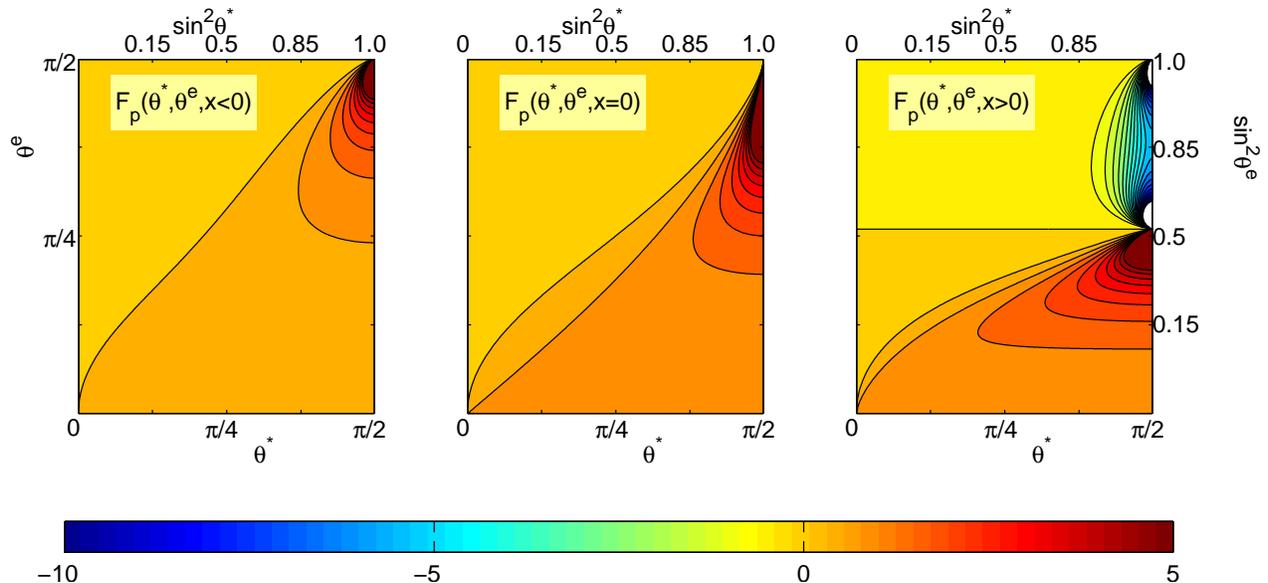}
  \end{center}
\caption{The function $F$ as defined in Eqn.~(\ref{FGH}) is plotted for three values of $X$: (a) $X=-0.001$ (left), (b) $X=0$ (middle) and (c) $X=0.001$ (right).
The central graph corresponds to $f_p$ since $X=0$.}
\label{mnalpha}
\end{figure}

Using new auxiliary functions, Eqn.~(\ref{non-K}) can be re-written as:
\begin{eqnarray}
\frac{6}{5}  f^{(4)}_{\rm NL}&=&
\left[
2J(\theta^*,\theta^e,X)\epsilon^*
-F(\theta^*,\theta^e,X)\eta^*_{\vp\vp}
-G(\theta^*,\theta^e,X)\eta^*_{\chi\chi}
\right.
%\nonumber
%\\
%&&
%\quad
+ {\rm sign}\left(b_\vp\right) {\rm sign}\left(\frac{U_\vp}{U}\right)
K(\theta^*,\theta^e,X)
\sqrt{\epsilon^*_b\epsilon^*_\vp}
\nonumber
\\
&&
\quad\left.
- 2 H(\theta^*,\theta^e,X)
\left(\hat\eta^e - \frac12{\rm sign}\left(b_\vp\right) {\rm sign}\left(\frac{U_\vp}{U}\right)
\frac{\sin^4\theta^e}{\cos\theta^e}\epsilon^e\sqrt{\frac{\epsilon_b^*}{\epsilon^*}}
-4\sin^2\theta^e\cos^2\theta^e\epsilon^e\right)
\right],
\end{eqnarray}
where
\begin{eqnarray}
F_p(\theta^*,\theta^e,X)&\equiv&
\frac{e^{-X}\alpha^3\tan^4\theta^*}{
\left(\alpha^2\tan^2\theta^*+\tan^4\theta^e\right)^2\cos^2\theta^e},
%\qquad
\nonumber\\
G_p(\theta^*,\theta^e,X)&\equiv&
\frac{e^{-X}\tan^8\theta^e}{\left(\alpha^2\tan^2\theta^*+\tan^4\theta^e\right)^2\sin^2\theta^e},
\nonumber\\
H_p(\theta^*,\theta^e,X)&\equiv&
\tan^2\theta^e\frac{\left(\alpha\tan^2\theta^*-\tan^2\theta^e\right)^2}{
\left(\alpha^2\tan^2\theta^*+\tan^4\theta^e\right)^2},
%\qquad
\nonumber \\
J_p(\theta^*,\theta^e,X)&\equiv&
e^{-X}\frac{\sin^2\theta^*}{\cos^2\theta^e}
\frac{\left(\alpha^3\tan^2\theta^*+\tan^6\theta^e\right)}{
\left(\alpha^2\tan^2\theta^*+\tan^4\theta^e\right)^2},
 \nonumber \\
K_p(\theta^*,\theta^e,X)&\equiv&
e^{-X}\alpha^2\frac{\cos^4\theta^e\sin^2\theta^e}{\sin^2\theta^*}.
%\nonumber
\label{FGH}
\end{eqnarray}
If $X=0$, then $\alpha=1$ and $F_p=f_p$, $G_p=g_p$, $H_p=h_p$ and $J_p=j_p$. The function $F_{p}$ is plotted for various values of $X$ in Figs.~\ref{mnalpha}.
From this definition, it is clear that allowing $\alpha\neq 0$, the range of $\theta^e$
for which we can obtain large $f_{NL}^{(4)}$ opens up. We also note that the symmetry between two fields is broken, due to
$b(\vp)$, which is apparent in Figs.~\ref{mnalpha}.
Note, though, that we require $\theta^*\sim0,\frac{\pi}{2}$ as before.
Furthermore, when $X>0$ (or $b_{e}>b_{*}$), we no longer require $\ces\ll1$.

Expressions for the power spectra and spectral index with a non canonical kinetic term are given in \cite{Choi:2007su,DiMarcoFinelli03}.

%%%%%%%%%%%%%%%%%%%%%%%%%%%%%%%%%%%%
\section{Conclusion}
\label{Section_conclusion}
We have made an in depth investigation into the level of
non-Gaussianity during two-field slow-roll inflation. We have shown that it is possible
to generate a large level of non-Gaussianity during inflation without violating slow roll and
when the inflaton field perturbations are Gaussian at Hubble exit.

The general conditions for generating a large non-Gaussianity show that the inflaton
potential must have a specific shape so that the angle of the background trajectory can
grow by about two orders of magnitude or more during inflation. In the case of an
inflation potential made of a product of two quadratic potentials, this condition is not
possible so that we conclude this model cannot generate a large non-Gaussianity during
slow roll inflation. When the angle of the background trajectory grows sufficiently (in
relative terms), we still need one of the fields to dominate throughout inflation, yet
the remaining field can not remain full insignificant. For typical values of the slow
roll parameters, we require that initially the subdominant field, say $\vp$, must satisfy
$\cis=\epspi/\epsilon^*\lesssim10^{-7}$. This means that the field trajectory is almost
exactly parallel to the $\chi$ axes initially and typically this requires a finely tuned
initial condition. In the case of a product separable potential we then also require that
the final value of the angle of the background trajectory lies in a narrow range much
greater than the initial value (in relative terms) but still nearly parallel to the
$\chi$ axis. The analysis for a sum separable potential is very similar but more
complicated to analyse. The initial background trajectory must be similarly fine-tuned to
be nearly parallel to one of the axes of the inflaton fields, but the tuning at the end
of inflation is not written so simply. We have also shown how the constraints on
generating a large non-Gaussianity may be eased if we generalise the inflaton model to
allow a non-canonical kinetic term.

We have presented two explicit models where a large non-Gaussianity can be generated
during slow roll inflation. One is a product potential driven by a field with a quadratic
potential, which ends inflation when it approaches the minimum of the potential. If the
second field with an exponential potential has a sufficiently small initial value then
for certain values of the model parameter, $\lambda$, this field grows by the right amount
to generate a large non-Gaussianity, $\fnl\sim100$. We also consider the sum separable
model of hybrid inflation. We provide a few specific choices of the model parameters,
which are effectively $\etap$ and $\etac$, for a suitably large ratio of the initial
field values this model generates a large non-Gaussianity. We find this is possible either
if the $\eta$'s have the opposite sign (so that inflation takes place near a saddle point) or
when both of the $\eta$'s have the same sign. However we have also found that if one of
the field values is too small then quantum fluctuations may perturb the background
trajectory to an extent that the classical trajectory in field space is no longer valid.
There is also the possibility that the large scale loop correction which we have not
generally considered in this paper may not be negligible. We intend to return to these
issues and make a more thorough investigation of the hybrid model in a future
publication.

%%%%%%%%%%%%%%%%%%%%%%%%%%%%%%%%%%%%%
\acknowledgements

The authors thank Carsten van de Bruck,  Filippo Vernizzi and David Wands for useful comments. CB
acknowledges financial support from the Deutsche Forschungsgemeinschaft. K.-Y.C. is supported by
the Ministerio de Educacion y Ciencia of Spain under Proyecto Nacional FPA2006-05423 and
by the Comunidad de Madrid under Proyecto HEPHACOS, Ayudas de I+D S-0505/ESP-0346.  LMHH
acknowledges support from STFC.

%\newpage

\end{document}